\documentclass[aip,reprint]{revtex4-1}

\usepackage{graphicx}

\begin{document}

%\title{Femtosecond laser ablation of a metal, a dielectric and a semiconductor illuminated at oblique angles of incidence}
%\title{Universal definition of femtosecond ablation threshold with oblique illumination}
\title{Universal threshold for femtosecond laser ablation with oblique illumination}

\author{Xiao-Long~Liu}
\affiliation{College of Optical Sciences, University of Arizona, Tucson, AZ 85721, USA}
\affiliation{Academy of Opto-Electronics, Chinese Academy of Science, Beijing 100094, China}

\author{Weibo~Cheng}
\affiliation{College of Optical Sciences, University of Arizona, Tucson, AZ 85721, USA}
%\affiliation{Apple, 1 Infinite Loop, Cupertino, CA 95014, USA}

\author{Massimo~Petrarca}
\affiliation{La Sapienza University, SBAI Department, via A. Scarpa 14, 00161 Rome, Italy}

\author{Pavel~Polynkin}
\email{ppolynkin@optics.arizona.edu}
\affiliation{College of Optical Sciences, University of Arizona, Tucson, AZ 85721, USA}

%\date{\today}
 
\begin{abstract}
We quantify the dependence of the single-shot ablation threshold on the angle of incidence and polarization of a femtosecond laser beam, 
for three dissimilar solid-state materials: a metal, a dielectric and a semiconductor. 
Using the constant, linear value of the index of refraction, we calculate the laser fluence transmitted through the air-material interface 
at the point of ablation threshold.
We show that, in spite of the highly nonlinear ionization dynamics involved in the ablation process, the so defined transmitted threshold fluence is universally independent of the angle of incidence and polarization of the laser beam for all three material types.  
We suggest that angular dependence of ablation threshold can be utilized for profiling fluence distributions in ultra-intense femtosecond laser beams.
\end{abstract}


%PACS: 79.20.Eb - Laser ablation; 42.25.Bs - Wave propagation, transmission and absorption; 
% 42.60.Jf 	Beam characteristics: profile, intensity, and power; spatial pattern formation

\maketitle

Studies of femtosecond laser ablation are motivated both by its applications in micromachining 
and by the need to understand optical damage by intense femtosecond laser pulses. Virtually all of the prior studies have been conducted 
for the cases in which the ablating laser pulse strikes the surface of the material at normal angle of incidence (AOI). 
However, in practice, many if not most optical elements are used at oblique AOIs (most commonly at 45$^{\circ}$ and at Brewster's angle). 
In spite of the widespread use of optics with oblique incidence of the laser beam, there has been no systematic study reported  
on the dependence of ablation (or damage) threshold on the AOI for different materials.

The few prior investigations of femtosecond laser ablation at not-normal AOIs were focused on specific applications.
In particular, the dependence of the rate of ablation on the AOI and polarization for copper has been reported \cite{metal}. 
An effective penetration depth of the laser field into copper, different from the penetration depth calculated classically 
using the complex index of refraction, has been introduced to model the observed measurements. 
Ablation thresholds for polyimide films, used as a model material for corneal tissue, has been quantified at various AOIs and polarizations \cite{polyimide}.

In this paper, we report experimental results on the dependence of laser ablation threshold on the AOI and polarization 
of the incident laser beam, for dissimilar solid-state materials: a metal, a dielectric and a semiconductor. We find that at large AOIs and for S-polarization,
metals can be as resistant to optical damage as dielectrics. 
We show that the laser fluence transmitted through the air-material interface, calculated as if the reflective and absorptive properties 
of the material were not affected by the ablating laser pulse, is remarkably independent of the AOI and polarization of the laser beam.
Consequently, transmitted laser fluence at the point of ablation threshold can serve as a universal parameter that defines ablation threshold with oblique illumination. Furthermore, the dependence of ablation threshold on the AOI can be used for profiling fluence distributions in ultra-intense laser beams.
 
Our experiments make use of a commercial Ti:Sapphire ultrafast laser system that delivers 60\,fs pulses with 4\,mJ of energy per pulse 
at a pulse repetition rate of 1\,kHz. The optical spectrum of the laser pulses is centered at 800\,nm. The energy of the laser pulses used
for ablation can be continuously varied
by a half-wave plate followed by a reflection off a glass wedge at Brewster's angle. 
The orientation of the linear polarization of the laser beam incident on the sample is controlled via another half-wave plate. 
In our experiments, we investigated the cases of P- and S-polarizations, which correspond to the direction of the electric field vector
of the laser beam parallel and perpendicular to the plane of incidence, respectively. 

The 9\,mm-diameter beam from the laser system is apertured down to 4\,mm diameter and focused on the sample's surface 
using a lens with a 50\,mm focal length. The resulting $1/e^{2}$ beam diameter in the focal plane of the lens is 18\,$\mu$m.
We calculated that the profile of the laser beam propagating linearly is very close to a Gaussian when it reaches the sample's surface.  
The sample is attached to a manual rotation stage with the angular accuracy of 0.1$^{\circ}$. 
The rotation stage is mounted on a computer-controlled motorized linear translation stage with the
maximum travel speed of 50\,mm/s, which enables ablation in the single-shot regime. 
Care is exercised in aligning the setup so that the front surface of the sample always stays in the focal plane of the focusing lens 
while the sample is translated by the motorized stage. 
For each of the AOIs that we investigated, the surface of the sample remains within the beam's Rayleigh range. 

\begin{figure}[b]
\includegraphics[width=8.6cm]{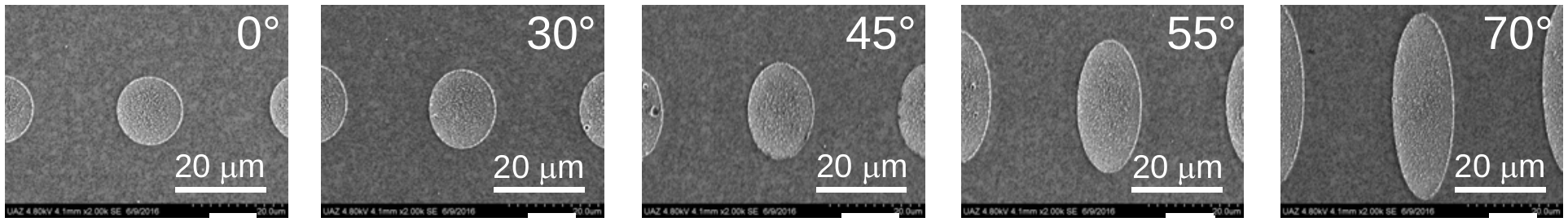}
\caption{SEM images of single-shot ablation craters produced on a gold surface by 
an S-polarized, 60\,fs\,--\,long laser pulse with 6.5\,$\mu$J of energy at 800\,nm wavelength, 
at different angles of incidence, as indicated in the individual images.}
\end{figure}

For every particular value of AOI set by the rotation stage, we produce single-shot ablation marks with increasing values of energy of the
ablating laser pulse. Ablation is performed in the ambient air. 
Shown in Figure~1 are example SEM images of the ablation craters for the case of a gold sample 
ablated by an S-polarized, 6.5\,$\mu$J pulse. The sample consists of a 1000\,\AA\,--\,thick gold layer 
deposited on a silicon wafer. The ablation craters have sharp and clearly defined boundaries, which indicates that femtosecond laser ablation is a threshold process
with respect to the local laser fluence. As expected, the ellipticity of the craters grows as $\propto$\,1/cos(AOI) as the AOI is increased.

In order to find the value of ablation threshold fluence at a particular AOI, we plot the lengths of the short 
and long axes of the elliptical ablation craters, squared, vs. the logarithm of the pulse energy. If the laser beam is Gaussian,
these plots show linear dependence, with the slope equal to twice
the $1/e^{2}$ beam radius squared; extrapolating the linear dependence of crater size out to zero yields the value
of ablation threshold fluence \cite{log}. 

For our experimental data to be consistent, the following four conditions need to be satisfied:
(i) The dependencies of the lengths of the long and short axes of the elliptical ablation craters squared,
on the logarithm of the pulse energy, are linear;
(ii) the size of the laser beam incident on the sample, extracted from the slope of the data curve for the short axis, 
is independent of the AOI and equal to the focused beam size, which is calculated assuming linear propagation; 
(iii) the size of the beam extracted from the data for the long axis, multiplied by the cos(AOI),
is also independent of the AOI and equal to the focused beam size, calculated assuming linear propagation; 
(iv) the values of the ablation threshold fluence obtained from the extrapolations of the data for the long and short axes, 
at a particular AOI, are equal to each other. All the data shown in this paper satisfy the above consistency criteria. 
Examples of the data obtained with a gold sample illuminated at 45$^{\circ}$ AOI, 
for the cases of P- and S-polarizations of the ablating laser beam, are shown in Figure~2.

We point out that these measurements need to be performed at sufficiently low values of pulse energy, so that the length of the short 
axis of the elliptical ablation crater does not significantly exceed the FWHM beam size of the incident beam. If the energy is too high 
and the corresponding length of the short axis significantly exceeds the beam size, the deviations of the spatial beam profile 
from the Gaussian shape, which are inevitably present at the periphery of the beam, will cause the dependence shown in Figure~2 to become nonlinear, 
thus increasing the uncertainty of the data extrapolation to zero crater size. 
We also point out that the peak fluence on the beam axis can significantly exceed the value of ablation threshold,
thus enabling the onset of thermal effects in ablation \cite{Nolte2} and material blow-off \cite{Bouilly}; this is of no importance here,
as we are not interested in the appearance of the center of the ablation crater, but only of its boundary, 
where the local laser fluence is always equal to the ablation threshold fluence.

Even though our experiments are conducted in air, and air ionization may occur for some of the values of pulse energy 
that we use, the generated plasma 
has negligible effect on the propagation of the laser beam towards the sample, because very tight focusing of the laser beam 
(with the corresponding f-number of 12.5) always dominates diffraction due to plasma.
Accordingly, for all of the data shown in this paper, the laser beam diameter extracted from the slopes of the ablation curves, 
such as shown in Figure~2, is equal to the value calculated assuming linear propagation of the focused laser beam. 

\begin{figure}[t]
\includegraphics[width=8.6cm]{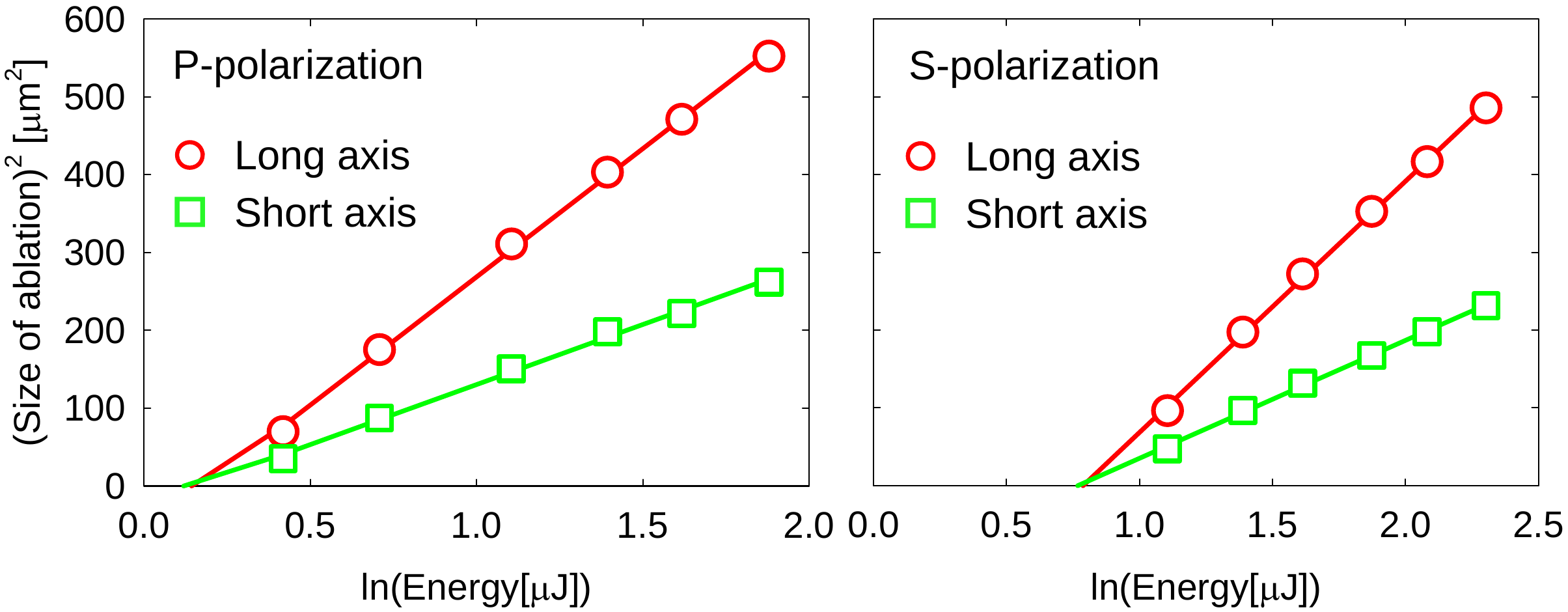}
\caption{Examples of the data for the size of the ablation crater squared vs. logarithm of the energy of the ablating laser pulse,
for the case of a gold sample illuminated at 45$^{\circ}$ AOI. The left and right panels show P- and S-polarizations of the incident laser beam, respectively. The two lines of data on each plot correspond to the long and short axes of the elliptical ablation crater. The experimental errors in these data
are small compared to the size of the data symbols.}
\end{figure}
 
Generally, the threshold for femtosecond laser ablation is reached when a specific amount of laser energy, per atom or molecule of the material, 
is absorbed in the top layer of the sample. This very general definition is independent of both the nature of the material and the specific ablation mechanism.
(The actual value of energy absorbed per atom or molecule at threshold does depend on the nature of the material and ablation mechanism,
but not on the AOI and polarization of the laser beam.) 
A rigorously computed value of fluence, absorbed per particle in the top layer of the sample, is a complex and nonlinear 
but definitely monotomically growing function of the incident laser fluence. Therefore it is natural to expect that ablation threshold fluence 
should be a growing function of the AOI, since the area of the sample that the incident fluence is spread over increases as $\propto$ cos(AOI).
In what follows, we will present data for ablation threshold fluence together with the corresponding values 
of threshold fluence multiplied by the factor $1/\mbox{cos (AOI)}$, as well as the approximate values of threshold fluence transmitted 
through the top surface of the material. Transmitted fluence will be calculated using Fresnel formulas with the constant (linear) 
complex index of refraction, as if the optical properties of the material were not perturbed by the incident laser field. 
Our results will show that transmitted fluence, as calculated above, at the point of ablation threshold is universally independent of both the AOI and polarization 
of the laser beam even though the complex index of refraction of those materials can change significantly 
within the duration of the ablating laser pulse \cite{Tinten}. We will discuss these findings after presenting our data.

\begin{figure}[b]
\includegraphics[width=8.6cm]{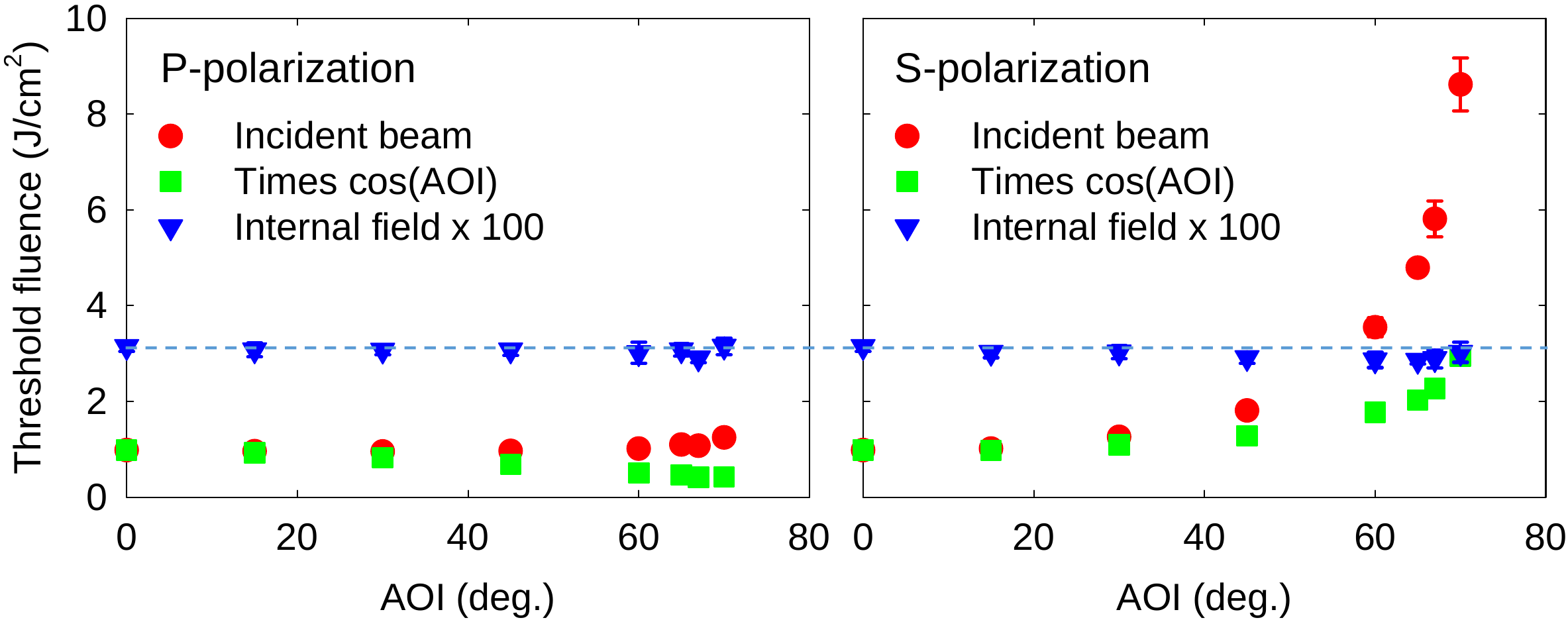}
\caption{Ablation threshold fluence and the quantities related to it, vs. AOI for gold. 
The internal fluence shown is multiplied by 100,
to make it visible on the plots. The experimental errors are due to the extrapolation uncertainty at large values of the AOI.}
\end{figure}

We start with the case of a metal. As in the cases shown in Figure 1 and 2, the sample is a 1000\,\AA\,--\,thick layer of gold deposited on a silicon wafer
(surface flatness $\lambda$/10). 
This film thickness is large enough for the gold layer to respond to the near-infrared femtosecond laser excitation 
as bulk gold \cite{Stewart}.
In Figure~3, we show data for the ablation threshold fluence, at different AOIs, together with the corresponding values of threshold fluence multiplied 
by the factor $1 / \mbox{cos (AOI)}$ and the corresponding transmitted fluence, computed with the unperturbed
complex index of refraction of gold $n \, = \, 0.1886 \, + \, i \cdot  4.7053$, at 800\,nm wavelength. 
In calculating the E-field immediately below the gold surface we used the standard Fresnel formulas with a complex refractive index.
For the case of a metal, the laser energy transmitted through the material surface is absorbed in the material entirely, therefore
in this case transmitted fluence is equivalent to absorbed fluence.

It is evident from the data that ablation threshold fluence weakly depends on the AOI for P-polarization and is a growing function of the AOI
for S-polarization. Threshold fluence times $\mbox{cos(AOI)}$ is a slightly decreasing function of the AOI for P-polarization 
and is a growing function of the AOI for S-polarization. 
However, the value of internal fluence at the point of ablation threshold is invariant with respect to both the AOI and polarization 
of the incident laser beam. 

The value of ablation threshold fluence for gold that we measure at normal incidence is (0.98\,$\pm$\,0.05)\,J/cm$^{2}$, which is close to the 
previously reported value \cite{Stewart}.
Note that throughout this paper, we show the peak value of ablation threshold fluence (on-axis fluence), 
which, for a Gaussian beam, is a factor of 2 larger than the average threshold fluence used by some authors.

\begin{figure}[t]
\includegraphics[width=8.6cm]{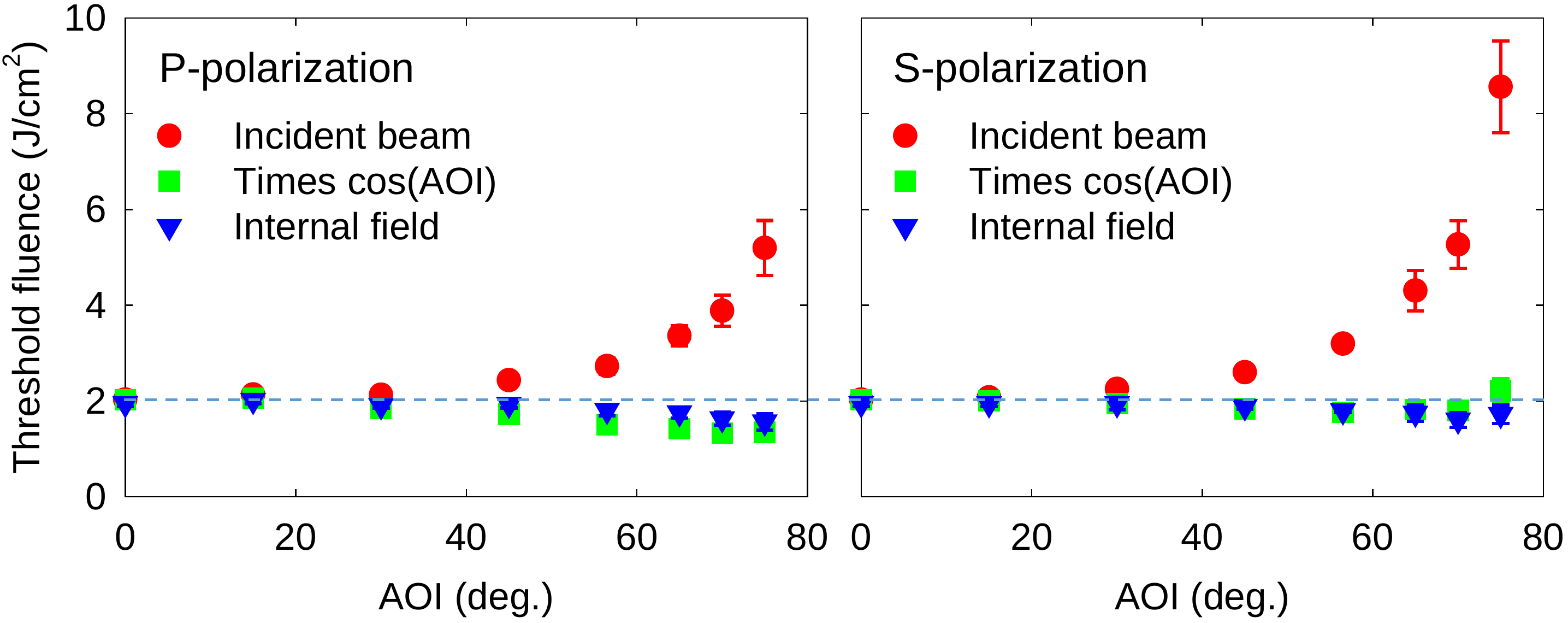}
\caption{Same as in Figure 3, but for the case of soda-lime glass.}
\end{figure} 

The second case we consider is that of a wide-bandgap dielectric material -- soda-lime glass. 
Our samples are standard microscope slides with surface flatness of $\lambda$/4.
Data for ablation threshold fluence, together with the related quantities, are shown in Figure~4. 
Here, the threshold fluence is a growing function of the AOI for both polarizations. 
Also for both polarizations, the threshold fluence multiplied by cos(AOI) is nearly independent of the AOI and so is the 
value of the transmitted fluence at the point of ablation threshold. 
The transmitted fluence is calculated
using a constant, unperturbed-by-the-laser-field, index of refraction of $n=1.45$, with a negligibly small imaginary part.  

The value of ablation threshold fluence for soda-lime glass that we measure at normal incidence is (2.03\,$\pm$\,0.07)\,J/cm$^{2}$. 
It is in good agreement with the value reported for borosilicate glass with 50\,fs pulses at 780\,nm wavelength \cite{Lenzner}.

The third material type we consider is a semiconductor. Our sample is a commercial, un-doped, single-crystal (100) silicon wafer
(the normal to the wafer's polished surface is parallel to the [100] direction in the crystal lattice). 
To illuminate the sample at an oblique AOI, we tilt it by rotating it around the [011] direction; the plane of incidence for the laser beam is perpendicular 
to this direction. 
Contrary to the previous cases of gold (polycrystalline) and glass (amorphous), silicon is a single-crystal, anisotropic material. 
Since silicon has a diamond-cubic crystal lattice, its optical properties are isotropic.
However, the anisotropy of the effective electron mass in silicon may result in a dependence of ionization and ablation on the orientation of the crystal,
as was previously shown for the case of multi-shot ablation \cite{SiAnisotropy}. 
The issue of ablation anisotropy that could result from the crystalline
nature of silicon is beyond the scope of this paper.  
In Figure~5, we show ablation threshold fluence, together with the related quantities,  
as functions of the AOI, for (100) silicon. As in the case of soda-lime glass discussed above, 
ablation threshold fluence is a growing function of the AOI for both polarizations. 
The threshold fluence times $\mbox{cos(AOI)}$ is also an increasing function 
of the AOI for both polarizations. 
However, as in the cases of metal and glass, the internal laser fluence at the point of ablation threshold 
is invariant with respect to the AOI and polarization. Here, the internal fluence is again calculated using a constant (unperturbed by the laser field) 
value of the complex index of refraction for silicon at 800\,nm wavelength, $n \, = \, 3.6959 + i \cdot 0.0047$.

The value of ablation threshold fluence for (100) silicon that we measure at normal incidence is (0.13\,$\pm$\,0.02)\,J/cm$^{2}$.
It is of the same order of magnitude as the value obtained through the extrapolation of the multi-shot threshold for the formation of nano-ripples on (100) silicon
to the number of laser shots equal to one \cite{multishotSi}. 
We are not aware of any previously reported data for single-shot ablation thresholds 
for (100) silicon with $\sim$50\,fs laser pulses at 800\,nm wavelength that can be directly compared with our results.

\begin{figure}[t]
\includegraphics[width=8.6cm]{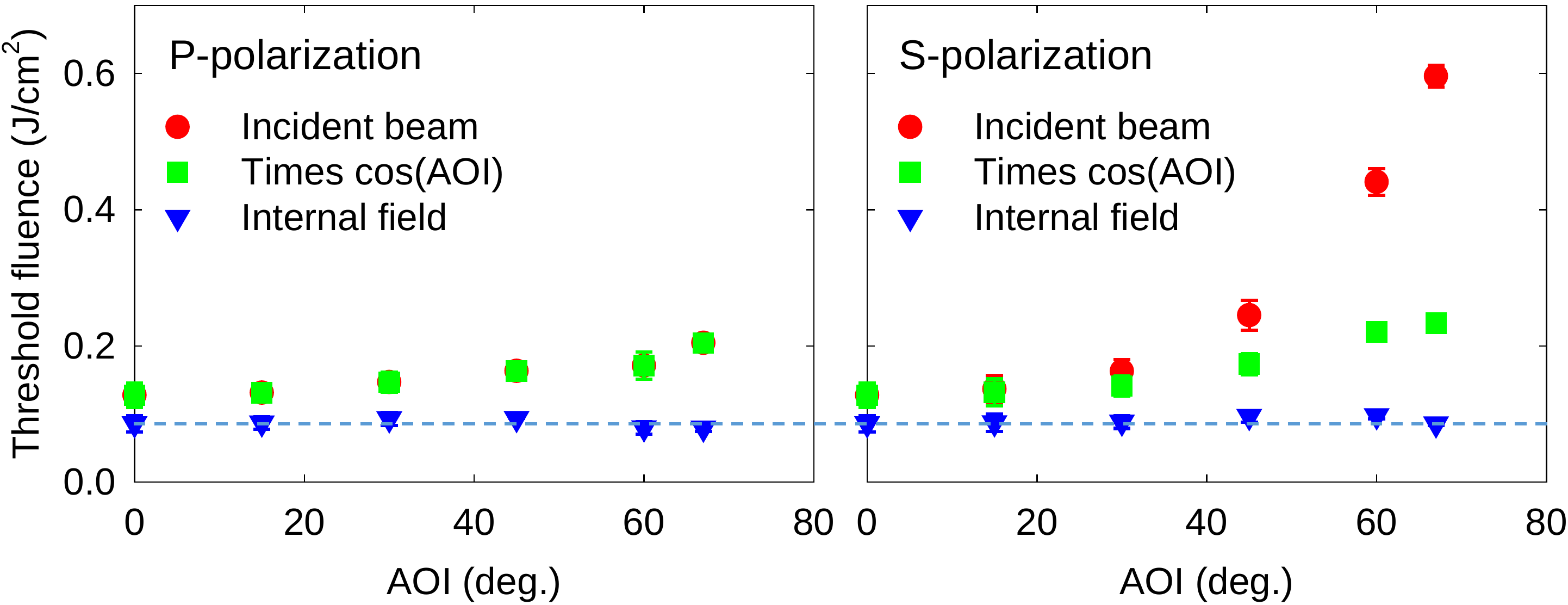}
\caption{Same as in Figures 3 and 4 but for the case of (100) silicon.}
\end{figure}

As pointed out above, the value of the absorbed laser energy per particle at ablation threshold should be independent of the AOI and polarization.
In general, fluence absorbed in the top layer of the sample is a complex and nonlinear function of the incident laser fluence, as both reflective and 
absorptive properties of the material are transiently altered by the ablating laser pulse. Our data empirically show that the quantity equal to the 
transmitted laser fluence, computed as if the reflective properties of the sample were unchanged through the interaction of the sample with the laser pulse, 
is nearly independent of the AOI and polarization. 
%That observation empirically suggests that thus calculated transmitted fluence and the total energy absorbed in the
%top layer of the sample, are proportional to each other, and the proportionality coefficient is independent of the AOI and polarization. 
For the case of gold,
such an observation is expected, as reflective and absorptive properties of gold weakly change during the interaction. In the cases of glass and silicon,
we argue that transmission of the incident laser energy into the sample decreases, while absorption in the top layer of the sample increases, 
as the sample is ionized by the laser pulse. Our data empirically show that the two effects compensate each other to a large degree.

A known dependence of ablation threshold fluence on the AOI can be used for the measurements of transverse fluence distributions 
in ultra-intense laser beams, such as femtosecond beams propagating in air in the self-channeling (filamentation) regime \cite{filaments}. 
For example, the data for gold illuminated with S-polarized light shown in Figure~3 suggest that gold can be used for profiling intense laser beams 
in the range of fluences from 1 to 9\,J/cm$^{2}$. For 50\,fs FWHM pulse duration, this range of fluence corresponds to the range of peak power 
from 20 to 170\,TW/cm$^{2}$. 

%To profile an intense laser beam, a material that has a quantified strong dependence of the ablation threshold 
%fluence on the AOI needs to be used. We would produce single-shot ablations of that material, at different AOIs, with the beam 
%that has the unknown fluence distribution we wish to quantify. The resulting elliptical ablation patterns would be scaled through the compression 
%along the long axis of the ellipse by the $\mbox{cos(AOI)}$ factor. The boundaries of the resulting scaled patterns 
%will outline the slices of the beam profile at different levels of fluence, equal to the values of the ablation threshold fluence at the AOIs 
%used in these beam-profiling experiments. Note that such a profiling procedure does not assume any particular beam shape or axial symmetry.

%Complex temporal pulse dynamics that may accompany highly nonlinear propagation of ultra-intense laser beams in air 
%\cite{filamentation} will not have a significant effect on the accuracy of the fluence (not intensity) profiling through the procedure discussed above, 
%because the value of ablation threshold
%fluence is weakly dependent on pulse duration, as long as the pulse is shorter than few hundred femtoseconds but longer than $\sim$\,30\,fs 
%\cite{Stewart,Kieffer}. The ultra-short temporal spikes that may develop in the intensity profile of the laser pulse 
%will also not significantly affect the accuracy of the fluence profiling
%because those spikes carry only a small fraction of the overall fluence of the laser beam.

In conclusion, we have experimentally investigated the dependence of threshold fluence for single-shot femtosecond laser ablation on the angle of incidence
and polarization of the ablating laser beam. We have found that for optically dissimilar materials (metal, glass, semiconductor), the value
of fluence transmitted into the material at the point of ablation threshold, if calculated using the linear (unperturbed by the laser field) 
complex index of refraction, is nearly independent of the angle of incidence and polarization.
%These findings suggest that ablation is driven by the fluence or by the electric field of the ablating laser pulse immediately below the material surface.
We suggest that the dependence of ablation threshold fluence on the angle of incidence can be utilized for profiling fluence distributions in ultra-intense laser beams
in the range from a fraction of 1\,J/cm$^{2}$ up to about 10\,J/cm$^{2}$.   

We thank Dr. Colm Dineen for helpful discussions and Mr. Lee Johnson and Prof. Thomas Milster for the assistance with operating the electron microscope.
This work was supported by the United States Air Force Office of Scientific Research under programs FA9550-12-1-0482 and FA9550-16-1-0013
and by the US Defense Threat Reduction Agency under program HDTRA 1-14-1-0009. 
X.-L.~Liu acknowledges the support from the National Natural Science Foundation of China under grants No. 11404335 and No. 91538113.


\begin{thebibliography}{99}
%\bibitem{Moureau} A.-C.~Tien, S.~Backus, H.~Kapteyn, M.~Murnane, G.~Mourou,
%``Short-pulse laser damage in transparent materials as a function of pulse duration",
%Phys. Rev. Lett. {\bf 82}, 3883 (1999).
%\bibitem{Nolte} B.~Chichkov, C.~Momma, S.~Nolte, F.~von~Alvensleben, A.~T\"{u}nnermann,
%``Femtosecond, picosecond and nanosecond laser ablation of solids",
%Appl. Phys. A {\bf 63}, 109 (1996).
%\bibitem{Mazur} R.~Gattass, E.~Mazur, 
%``Femtosecond laser micromachining in transparent materials",
%Nat. Photonics {\bf 2}, 219 (2008). 
\bibitem{metal} Y.~Miyasaka, M.~Hashida, T.~Nishii, S.~Inoue, S.~Sakabe,
Appl. Phys. Lett. {\bf 106}, 013101 (2015).
\bibitem{polyimide} B.~S.~Haq, H.~U.~Khan, K.~Alam, M~Mateenullah, S.~Attaulah, I.~Zari,
Appl. Opt. {\bf 54}, 7413 (2015).
\bibitem{log} J.~M.~Liu, Opt. Lett. {\bf 7}, 196 (1982).
\bibitem{Nolte2} S.~Nolte, C.~Momma, H.~Jackobs, A.~T\"unnermann, B.~N.~Chichkov, B.~Wellegehausen, H.~Welling,
J. Opt. Soc. Am. B {\bf 2716} (1997).
%\bibitem{layer} T.~Kumada, T.~Otobe, M.~Nishikino, N.~Hasegawa, T.~Hayashi,
%``Dynamics of spallation during femtosecond laser ablation studied by time-resolved reflectivity with double pump pulses",
%Appl. Phys. Lett. {\bf 108}, 011102 (2016).
\bibitem{Bouilly} D.~Bouilly, D.~Perez, L.~J.~Lewis,
%``Damage in materials following ablation byultrashort laser pulses: A molecular-dynamics study",
Phys. Rev. B {\bf 76}, 184119 (2007).
\bibitem{Tinten} K.~Sokolowski-Tinten, J.~Bialkowski, M.~Boing, A.~Cavalleri, D.~von~der~Linde,
%``Thermal and nontermal melting of gallium arsenide after femtosecond laser excitation",
Phys. Rev. B {\bf 58}, R11805 (1998).
\bibitem{Stewart} B.~Stewart, M.~Feit, S.~Herman, A.~Rubenchik, B.~Shore, M.~Perry,
%``Optical ablation by high-power short-pulse lasers",
J. Opt. Soc. Am. B {\bf 13}, 459 (1996).
%\bibitem{Byer} A.~Ben-Yakar, R.~Byer,
%``Femtosecond laser ablation properties of borosilicate glass",
%J. Appl. Phys. {\bf 96}, 5316 (2004).
\bibitem{Lenzner} W.~Kautek, J.~Kr\"uger, M.~Lenzner, S.~Sartania, C.~Spielmann, F.~Krausz,
%``Laser ablation of dielectrics with pulse durations between 20\,fs and 3\,ps",
Appl. Phys. Lett. {\bf 69}, 3146 (1996).
\bibitem{SiAnisotropy} X.~Li, W.~Rong, L.~Jiang, K.~Zhang, C.~Li, Q.~Cao, G.~Zhang, Y.~Lu,
%``Generation and elimination of polarization-dependent ablation of cubic crystals by femtosecond laser radiation",
Opt. Exp. {\bf 22}, 30170 (2014).
\bibitem{multushotSi} S.~He, J.~Nivas, A.~Vecchione, M.~Hu, S.~Amoruso, 
%"On the generation of grooves on crystalline silicon irradiated by femtosecond laser pulses",
Opt. Exp. {\bf 24}, 3238 (2016).
\bibitem{multishotSi} S.~He, J.~J.~J.~Nivas, A.~Vecchione, M.~Hu, S.~Amoruso,
%"ON the generation of grooves on crystalline silicon irradiated by femtosecond laser pulses",
Opt. Exp. {\bf 24}, 3238 (2016).
\bibitem{filaments} A.~Couairon, A.~Mysyrowicz, 
%"Femtosecond filamentation in transparent media",
Phys. Rep. {\bf 441}, 47
%-189
(2007).
%\bibitem{filamentation} M.~Gaarde, A.~Couairon,
%``Intensity spikes in laser filamentation: Diagnostics and applications",
%Phys. Rev. Lett. {\bf 103}, 043901 (2009).
%\bibitem{Kieffer} B.~Chimier, O.~Ut\'{e}za, N.~Sanner, M.~Sentis, T.~Itina, P.~Lassonde, F.~L\'{e}gar\'{e}, F.~Vidal, J.~C.~Kieffer,
%``Damage and ablation thresholds of fused-silica in femtosecond regime",
%Phys. Rev. B {\bf 84}, 094104 (2011).
\end{thebibliography}
\end{document}